# Ultrafast zero balance of the oscillator-strength sum rule in graphene

**Jaeseok Kim[1†], Seong Chu Lim[2,3†], Seung Jin Chae[2,3], Inhee Maeng[1], Younghwan Choi[1], Soonyoung Cha[1], Young Hee Lee[2,3*] & Hyunyong Choi[1*]**

[1] *School of Electrical and Electronic Engineering, Yonsei University, Seoul 120-749, Korea*

[2] *IBS Center for Integrated Nanostructure Physics, Institute for Basic Science, Sungkyunkwan University, Suwon, 440-746, Korea*

[3] *Department of Energy Science, Department of Physics, Sungkyunkwan University, Suwon 440-746, Korea*

[*]Correspondence and requests for materials should be addressed to Y.H.L. (leeyoung@skku.edu) or H.C. (hychoi@yonsei.ac.kr)

[†]These authors contribute equally to this work

**Oscillator-strength sum rule in light-induced transitions is one general form of quantum-mechanical identities. Although this sum rule is well established in equilibrium photo-physics, an experimental corroboration for the validation of the sum rule in a nonequilibrium regime has been a long-standing unexplored question. The simple band structure of graphene is an ideal system for investigating this question due to the linear Dirac-like energy dispersion. Here, we employed both ultrafast terahertz and optical spectroscopy to directly monitor the transient oscillator-strength balancing between quasi-free low-energy oscillators and high-energy Fermi-edge ones. Upon photo-excitation of hot Dirac fermions, we observed that the ultrafast depletion of high-energy oscillators precisely complements the increased terahertz absorption oscillators. Our results may provide an experimental priori to understand, for example, the**

**intrinsic free-carrier dynamics to the high-energy photo-excitation, responsible for optoelectronic operation such as graphene-based phototransistor or solar-energy harvesting devices.**

The power of the oscillator-strength sum rule lies in the rich physical information it contains concerning the fundamental characteristics of the energy spectrum, often in an experimentally amenable configuration. A primary feature of this spectrum is that the transition strength in a given system should be conserved during any optical or electronic excitations[1-3]. The sum rule, therefore, has been used in numerous areas of experimental physics, including atomic[2,4], molecular[2], particle[5], and nuclear physics[3], to substantiate a consistency check of a theory. One of well-established examples is an observation of a superconductivity gap, in which the validity of sum rule was estimated by studying the reduced sum-rule integral in the real-part of optical conductivity $\sigma_1(\omega)$ [6]. In many cases, however, application of the sum rule has remained elusive, requiring certain approximation in the measurable frequency range.

In conventional solids, the energy-dependent dispersion contains complicated absorption features associated with multiple optical resonances between/within the conduction and valence bands; thus, the sum rule is often of limited use even in the equilibrium regime. Graphene, however, provides an ideal model system for studying the intrinsic nature of sum-rule dynamics because its electronic dispersion closely resembles that of linear Dirac quasiparticles[7,8]. As illustrated in Fig. 1a, the graphene sum-rule relationship in equilibrium state implies that the integrated low-energy absorption oscillators in doped graphene are the same as the depleted interband oscillators below twice the Fermi level $|E_F|$[9], i.e.,

$$\int_0^{2|E_F|} \sigma_1(E)dE = \sigma_0 2|E_F|. \tag{1}$$

where $\sigma_1(E)$ is the real part of optical conductivity and $\sigma_0 = q^2\pi/2h$ is the universal quantum conductivity ($q$ is the electron charge and $h$ is Planck's constant)[7]. This relationship is also equivalent to the transfer rule of Drude-spectral weight. Equation (1) is valid within Drude approximation for the conductivity of graphene, in which the level of doping $|E_F|$ is assumed to be large compared with the scattering rate[9,10].

The central aspect to address in this paper is to secure whether the *time-dependent* sum rule still holds when the graphene is driven into *out-of-equilibrium*. When photo-excitation occurs across from the valence band to the conduction band (Fig. 1b), the excitation can invoke two divided transition oscillators; low-energy oscillators for intraband transition and high energy oscillators for interband transition near the $2|E_F|$. In this case, the two different transition strengths are supposed to balance each other such that the following simple relation holds[11-13]:

$$\int_0^\infty \Delta\sigma_1(\Delta t, E)dE = 0, \qquad (2)$$

where $\Delta\sigma_1(\Delta t, E)$ is the time-dependent change in the real part of the optical conductivity. This prediction has inspired us to use both ultrafast low-energy terahertz (THz) and high-energy optical spectroscopy to rigorously investigate the time- and spectral-dependent sum rule dynamics in an out-of-equilibrium regime. Although recent studies concerning the ultrafast characteristics of graphene have been published[14-16], to date no studies that involve *concurrent* probing of the *transient* oscillator dynamics in the THz region and the optical domain have appeared in the literature.

For practical application such as graphene photodetectors and hot carrier solar cells, the dynamic sum rule is important since the coupled sum-rule dynamics in THz and optical range can be useful to estimate the genuine magnitude of low-energy free-carrier dynamics (which is related to the nonequilibrium THz oscillators) when the photo-excited hot-carrier generation is required (that is related to the nonequilibrium high-energy optical oscillators)[17-20]. Furthermore, unlike conventional semiconductors exhibiting complicated intra- and interband absorptions, the pure linear dispersion of graphene imposes a fundamental physical restriction on the optical-to-electrical photoresponse, namely the *time-dependent* changes of the two oscillators are strongly coupled and should be balanced via Eq. (2).

The single-layer graphene used in this experiment was prepared via thermal chemical vapor deposition (CVD)[21,22] and was transferred onto an optically transparent CVD diamond substrate (see Methods for the sample preparation). Four-point-probe and Hall measurements were performed to determine the doping level of our sample ($E_F = -306$ meV; see Supplementary Information). For the ultrafast spectroscopy measurements, ultrashort 50 fs, 1.55 eV pulses from a 250 kHz Ti:sapphire regenerative amplifier system (Coherent RegA 9050) were used as a pump pulse to excite the sample. For the time-resolved, low-energy THz spectroscopy, broadband THz pulses covering the photon range of 7-18 meV were generated and detected via optical rectification and electro-optic sampling in a 0.3 mm thick, (110)-oriented GaP crystal. For the high-energy optical spectroscopy, tunable infrared pulses were generated through optical parametric amplification (Coherent OPA 9850) and used as interband probe photons. Because the linear dispersion of Dirac fermion breaks down as the photon energy approaches the nearest-neighbor hopping energy (~ 3 eV)[23], we have restricted the pump-photon energy of 1.55 eV in our experiment. The graphene sample was stored in a vacuum cryostat with CVD diamond windows, and all measurements were performed at liquid-nitrogen temperature of 77 K.

# Results

**Transient dynamics of low-energy oscillators in the THz range.** We first discuss the measurements on the time-dependent oscillator dynamics in the low-energy THz range. Picosecond THz pulses probe the pump-induced THz field change $\Delta E(t)$ as a function of the pump-probe delay $\Delta t$. Typical time-domain THz signals are presented in Fig. 2a. Non-zero signal changes with opposite signs between $\Delta E(t)$ and the reference THz field $E_0(t)$ are the key indications of the *transient absorption* of the THz oscillators[24,25]. In general, there are two contributions to the dynamic optical conductivity; minority carriers (electrons) and majority carriers (holes). By performing fluence-dependent studies, the low-energy absorption response can be demonstrated to increase following the relationship[25]

$$\text{Electron}: -\frac{\Delta E}{E_0} \propto \frac{\Delta \sigma}{\sigma_0} = \frac{4k_B T_C \ln(1+e^{\Delta E_{F,e}/k_B T_C})}{\pi \hbar} \frac{\tau}{1-i\omega\tau}, \quad (3)$$

$$\text{Hole}: -\frac{\Delta E}{E_0} \propto \frac{\Delta \sigma}{\sigma_0} = \frac{4\Delta E_{F,h}}{\pi \hbar} \frac{\tau}{1-i\omega\tau}, \quad (4)$$

where $k_B$ is the Boltzmann constant, $T_C$ is the hot-carrier temperature, $\Delta E_{F,e(h)}$ is the photo-induced electron (hole) quasi-Fermi level change, and $\tau$ is the momentum scattering time. Because the change of quasi-Fermi level for electrons $\Delta E_{F,e} = \hbar v_F \sqrt{\pi \Delta n}$ is large compared to that of hole quasi-Fermi level $\Delta E_{F,h} = \hbar v_F \sqrt{\pi} \Delta p / 2\sqrt{P_0}$, the nonlinear $\Delta \sigma$ of electrons contributes more to the measured conductivity change than the linear $\Delta \sigma$ of holes; here, $v_F$ is the graphene Fermi velocity, $\Delta n (= \Delta p) = 0.63 \times 10^{12}$ cm$^{-2}$ is the photo-excited electron (hole) density at the peak $\Delta E(t)$, and $P_0 = 6.88 \times 10^{12}$ cm$^{-2}$ is the hole-doping density (Supplementary Information). Figure 2b illustrates the characteristic nonlinear dependence of the peak $\Delta E(t)$, confirming a distinguished aspect of Dirac quasiparticles.

More insights can be obtained by inspecting the energy-dependent complex optical sheet conductivity $\Delta\sigma(\omega)$ [26] in Fig. 2c. We can faithfully fit both $\Delta\sigma_1(\omega)$ and $\Delta\sigma_2(\omega)$ at different $\Delta t$ without significantly changing the momentum scattering rate (Supplementary Information). This result implies that the pump-induced scattering enhancement plays a minor role in our case[27,28]. Immediately after photo-excitation, the frequency-dependent $\Delta\sigma_1(\omega)$ exhibits an increased broadband conductivity $\Delta\sigma_1(\omega) > 0$. The carrier response arises quasi-instantaneously, which agrees well with recent investigations on the rapid carrier thermalization occurring via elastic carrier–carrier scatterings[14,29]. At $\Delta t > 0.2$ ps, the spectral shape of $\Delta\sigma_1(\omega)$ closely follows a featureless Drude-like theory[30], and the thermalized hot carriers relax through Auger and phonon-induced recombination[24] with a mono-exponent decay of 1.2 ps.

The pump-induced THz field change $\Delta E(t)$ and the corresponding $\Delta\sigma_1(\omega)$ reflect both minority and majority carrier dynamics, above and below the Dirac point, respectively[25]. Opposing to the mirror-like symmetry in a band structure of grpahene, two different carriers exhibit somewhat different dynamics. In doped graphene, the THz response from the majority carrier of holes is smaller than that from the minority carrier of electrons, and its Drude amplitude, i.e., the spectral magnitude near the Fermi-edge, is proportional to $\hbar v_F \sqrt{\pi}\Delta p / 2\sqrt{P_0}$ following Eq. (4). In contrast, the minority carrier exhibits a nonlinear dependence that follows Eq. (3). The sum of transient THz absorption is governed by both the spectrally-integrated majority and the minority carrier because the pump-induced $\Delta\sigma_1(\omega)$ represents the average electron and hole density distributed over each quasi-Fermi surface[9,10]. Thus, we can understand the Drude-spectral weight[30], i.e., $\Delta\sigma_1(\omega)$ integrated over the photon energy, as being the total sum of low-energy oscillators from the majority and minority carriers described by Eq. (3) and Eq. (4), respectively. In addition, the Drude-spectral weight is directly related to the depleted carrier density near $2|E_F|$ by Eq. (1). This finding denotes that the *increased* absorption in the THz range are related to the straightforward interpretation of the *decreased* absorption near the $2|E_F|$. As discussed below, we clearly observe this phenomenon—specifically, that the increased non-vanishing integral sum of the low-energy absorption is directly accompanied by reduced absorption strength near the Fermi edge.

**Transient dynamics of high-energy oscillators near the Fermi level.** We now discuss the time-resolved dynamics of the high-energy oscillators near the Fermi edge. Figure 3a depicts the ultrafast response near the Fermi-edge oscillators with the same excitation pump fluence as that in the THz case. The photo-induced optical conductivity at $\Delta t \sim 0.2$ ps demonstrates the broadband depletion of the interband transition, resulting in $\Delta\sigma_1(\omega) < 0$ above 580 meV [14]. At delays longer than 0.2 ps, the optical conductivity rapidly recovers due to the reduced carrier occupation above $|E_F|+|\Delta E_{F,h}|$ and slightly increases below $|E_F|+|\Delta E_{F,h}|$ due to the cooling of hot Dirac fermions[29]. The blue shift of zero-crossing in the spectra (arrow mark), is an indication of the quasi-Fermi edge, whose time-dependent change is a direct signature of cooling kinetics of the occupation probability[29]. This is evident when the equilibrium and nonequilibrium distribution are compared (Fig. 3b). Here, the transient zero-crossings of the nonequilibrium interband transition can be readily explained by the temperature-dependent occupation probability via rapid hot-carrier thermalization and cooling dynamics[30].

For more detailed analysis, Fig. 3c presents the bi-exponential kinetics of hot-carrier dynamics[14,31] fitted with two time constants of 0.3 ps and 5.5 ps to the extracted hot-carrier temperature in Fig. 3a. Recent experimental and theoretical investigations reveal that a strongly coupled phonon results in extremely fast cooling dynamics of hot Dirac quasiparticles for the first 0.25 ps; later, slow relaxations of hot carriers are dominated by relatively slow acoustic-phonon scattering[15,29], which are clearly reproduced in our measurements with two decaying components of 0.3 ps and 5.5 ps, respectively[31]. Figure 3d demonstrates a linear increase of the peak transmission changes with increasing pump fluence. This observation results from the fact that in the high-energy absorption, $\Delta\sigma_1(\omega)$ is proportional to the quasi-Fermi level changes of holes $\Delta E_{F,h}$, which, in turn, linearly depend on the excited hole density $\Delta p$ via $\Delta E_{F,h} = \hbar v \sqrt{\pi} \Delta p / 2\sqrt{P_0}$.

We note that the largest reduction of optical conductivity by electrons at 645 meV in $\Delta t \sim 0.2$ ps reaches about 71% of the universal quantum conductivity $\sigma_0$ (Fig. 3a). This implies that the photo-induced population changes from the interband transition at a *finite* frequency do not fully represent the overall temporal kinetics of the oscillator transfer. Similarly, the maximum THz conductivity oscillator exhibits an increase in $\sigma_0$ as large as 770 % at 7 meV when $\Delta t \sim 0.2$ ps (Fig. 2c), which far exceeds the largest depletion in the high-energy oscillator. Thus, we find that the sum-rule of transient optical conductivity is proportional to the energy-integrated $\Delta\sigma_1(\omega)$ over the involved photon energy.

The key result of this paper is presented in Fig. 4. A plot of numerical fits to the measured integrals of $\Delta\sigma_1(\omega)$ as a function of $\Delta t$ explains that the decrease of the interband oscillator (blue squares) is directly accompanied by increased oscillators in the THz range (red squares) within the model-based fits (gray squares, see the black solid fits in Fig. 2c and Fig. 3a). This finding highlights that the sum of these two $\Delta\sigma_1(\omega)$ results in a zero sum of the transient $\Delta\sigma_1(\omega,\Delta t)$ and provides solid evidence that the transient oscillator strengths are precisely balanced even when the carrier distribution is driven into a nonequilibrium regime. Because no significant spectral reshaping is observed in the high-energy interband oscillators at $\Delta t < 1$ ps (see Fig. 3a), the low-energy THz oscillators are expected to show no distinctive spectral structure within the THz temporal resolution of 1 ps.

**Discussion**

Our ultrafast investigation demonstrated that the ensuring transient sum rule is precisely fulfilled when the graphene system is driven into an out-of-equilibrium state for the time scale of around a few ps. Of course, if the transient dynamics enter into the coherent interaction regime, i.e. time scale shorter than the dephasing or thermalization time (typically less than 10 fs in graphene), the sum rule may break down depending on the Rabi oscillation frequency and lead to intriguing phenomena of coherent population oscillation. The sum-rule dynamics underscores that high-energy photon injection should accompany direct photo-conductivity transfer to the low-energy free-carrier response. Thus, our investigation provides not only fundamental insights into the nonequilibrium oscillator dynamics of graphene, but more importantly accounts for an experimental guide for understanding the photo-induced electrical response of graphene optoelectronics such as solar-energy-harvesting photovoltaic[20] or photothermoelectric devices[18]. Alternatively, application of the sum-rule dynamics into the doped graphene or gate-biased graphene may lead to new hybridized phototransistor devices, which contain both the graphene contact (responsible for the free-carrier electrical response) and the photo-sensitive channels (responsible for the energy transfer from the photo-injected hot carriers)[20,32,33].

# Methods

**Sample preparation.** Copper foil with a thickness of 100 µm and 99.96% purity was loaded into a furnace and annealed at 1060°C for 2 hours under 1000 sccm of Ar gas and 200 sccm of $H_2$ gas. The copper foil was subsequently polished chemically and mechanically. The copper foil was moved into the furnace for the synthesis of single-layer graphene. The temperature was maintained at 1060°C, and then 1000 sccm of Ar, 10 sccm of $H_2$, and 3 sccm of $CH_4$ were introduced into the chamber. The synthesis was complete in 2 minutes. To transfer the CVD-grown graphene onto a diamond substrate, we etched the copper foil using a Cu etchant (CE-100, Transene) and subsequently rinsed it with DI water several times. To protect the graphene during the transfer process, the graphene layer was covered with PMMA using a spin coater. After the graphene was transferred, we removed the PMMA using acetone.

**Experimental setup of ultrafast THz and optical spectroscopy.** To investigate the low-energy oscillator dynamics, ultrafast optical-pump THz-probe spectroscopy was performed. Ultrashort, 50 fs, 1.55 eV pump pulses were used to excite the sample in both the THz and optical experiments. A regenerative amplifier system (based on a 250 kHz repetition rate, Coherent RegA 9050) was optimized to accommodate the trade-off between the pump fluence and a high signal-to-noise ratio (SNR). For the time-resolved THz spectroscopy, broadband THz pulses covering the photon range of 7-18 meV were generated by optical rectification in 0.3 mm thick, (110)-oriented GaP crystals under dry air conditions. A set of a half-wave plate and polarizer was used to control the polarization and the power of generation, detection, and excitation pump pulses and to optimize the generated THz pulses. For the THz detection, an electro-optic sampling technique was used in the same 0.3 mm thick, (110)-oriented GaP crystal as for the THz generation. During the electro-optic sampling, the THz pulse induces birefringence; hence, the polarization of the co-propagating sampling pulse is altered to be slightly elliptical. A quarter-wave plate causes a phase shift of λ/4 in the incident pulse, and the sampling pulse then becomes more elliptical compared with the circular polarization without the induced THz pulse. A Wollaston prism was used to split the elliptically polarized sampling pulse into two orthogonal components, which were sent to a pair of balance detectors. The detector measures the intensity difference between the two orthogonal components, which is proportional to the applied THz amplitude. To study the high-energy interband oscillator dynamics, ultrafast optical-pump optical-probe spectroscopy was performed. In the optical parametric amplifier system (Coherent OPA 9850), the pump pulse with 50 fs pulse width and 1.55 eV center photon energy was used as a "seed" to generate a white-light continuum (0.4 - 1.1 µm) in a sapphire crystal. The white-light continuum was then combined with the split pump pulse to generate wavelength-tunable pulses of signal (1.2 - 1.4 µm) and idler (1.8 – 2.4 µm) in a beta-barium borate (BBO) crystal. We used the idler pulse with 612 meV center energy to investigate the Fermi-edge oscillators in our graphene sample.

**Phase check of the pump-induced THz field change.** We have carefully verified the relative phase between $\Delta E(t)$ and $E_0(t)$ using the following approach. First, the pump pulse was detected by a large-area photo-diode (Thorlabs DET100A-Si Detector, diameter of 9.8 mm) before exciting the sample. After the auto phase on a lock-in amplifier (Standard Research Systems SR 850) was set, the phase was read to be $\theta_0$. The phase of the peak $\Delta E(t)$ at $\Delta t = 0.2$ ps was checked using the same chopper and the same lock-in amplifier to read the phase to be $\theta_0 + \alpha$. Second, the 1.55-eV, 50 fs pulses that generated the THz pulses were chopped, and the photo-diode signal was measured using the same large-area photo-diode used to read the phase of $\theta_1$. Then, the phase of $E_0(t)$ was measured to be $\theta_1 + \beta$. Notably, the phase difference $(\theta_0 + \alpha) - (\theta_1 + \beta)$ results in $180 \pm 3°$. The 180° phase angle indicates that $\Delta E(t)$ has a sign opposite that of $E_0(t)$, which implies that our transient THz dynamics of $\Delta\sigma(\omega, \Delta t)$ represent an absorption oscillator. The high-energy optical measurement was verified using the same approach.

**Figure Legends**

**Figure 1. Schematics representing oscillator-strength sum rule.** (a) Energy-dependent linear dispersion in graphene and the corresponding absorption spectrum for the doped (yellow) and undoped (green) case. When graphene is doped, the flat absorption oscillators below $2|E_F|$ are transferred to the low-energy oscillators (black arrow). The absence of any complicated absorption spectra is a particular characteristic of graphene. (b) The schematic kinetics of the oscillator transfer from the high-energy interband to the low-energy intraband for the doped graphene. The equilibrium absorption (yellow) and the nonequilibrium absorption transfer (blue) are displayed.

**Figure 2. Transient dynamics of the low-energy THz oscillators.** (a) Negative THz field changes $-\Delta E(t)/E_0$ at different $\Delta t$ are displayed as a function of the field delay $t$ when the excited pump fluence is 40 µJ/cm². (b) Pump-induced maximum changes in $-\Delta E(t)/E_0$ are shown as a function of the pump fluence. (c) Pump-induced real and imaginary conductivity changes $\Delta\sigma(\omega)/\sigma_0$ (red circles) are plotted as a function of $\Delta t$. For analysis, we used the thin-film conductivity formula of $\Delta\sigma = -1/Z_0(\Delta E/E_0)(n_S+1)$, where $Z_0$ is the vacuum impedance, and $n_S = 2.4$ is the CVD diamond substrate refractive index. The measured conductivities were fitted using a Drude-like free-carrier model (black lines; see Supplementary Information).

**Figure 3. Transient dynamics of the high-energy oscillators near the Fermi level.** (a) Depletion dynamics of the high-energy oscillator are shown for several $\Delta t$. The transient $\Delta\sigma_1(\omega)/\sigma_0$ is obtained from the measured transmission change $\Delta T/T_0$ via the thin-film

formula of $\Delta\sigma_1 = 1/Z_0[\{1/(1+\Delta T/T_0)\}^{1/2} - 1](n_S + 1)$. Because $\Delta\sigma_1(\omega)$ is directly proportional to the occupation probability, we use a simple Fermi-distribution function to fit all the measured data (black lines). The zero-crossing in the measured conductivity exhibits a strong blue-shift (black arrow). (b) The equilibrium (black solid) and nonequilibrium conductivity spectra are presented for several $\Delta t$ values. The temperature broadening is a fit result to the $\Delta\sigma_1(\omega)$ spectra of Fig. 3a. (c) The extracted carrier temperature of the hot-Fermi distribution (blue squares) in Fig. 3a is well explained by bi-exponential decay kinetics (black solid line) with two decay time constants of 0.3 ps and 5.5 ps. Black dashed lines are guides to represent the two decaying components. (d) The measured transmission change (blue square) $\Delta T/T_0$ at 600 meV exhibits a linear dependence on the pump fluence.

**Figure 4. Dynamic zero-balance of the integrated absorption oscillators.** The depleted absorption near $2|E_F|$ is precisely balanced by the increased absorption in the low-energy THz data. The filled squares are the spectrally integrated sum of the THz oscillators (red) and the $2|E_F|$ oscillators (blue). The empty gray squares represent the sum of the two oscillators. Note that the zero balance of the dynamic sum rule is fulfilled within the temporal resolution of THz pulse of around 1 ps.

**Acknowledgements**

The work at Yonsei was supported by the Basic Research Program through the National Research Foundation of Korea (NRF) funded by the Ministry of Education, Science and Technology (No. 2011-0013255), the NRF grant funded by the Korean government (MEST) (NRF-2011-220-D00052, No. 2011-0028594, No. 2011-0032019) and the LG Display Academic Industrial Cooperation Program. S. C. Lim and Y. H. Lee at SKKU are grateful for the support from the Research Center Program of IBS.

**Author contributions**

J.K. and H.C. contributed to the experimental idea. J.K. and S.C.L. designed and carried out the experiments. J.K. and Y.C. developed theoretical modelling and performed simulations. S.C.L. and S.J.C. prepared and characterized the sample. J.K., S.C.L., S.J.C., I.M., Y.C., S.C., Y.H.L, and H.C. contributed to interpretation of the measured and analysed results. Y.H.L and H.C. initiated the work, managed the project, and interpreted the data. All authors discussed the results and commented on the manuscript.

**Additional information**

**Supplementary information** accompanies this paper at http://www.nature.com/Scientificreports

**Competing financial interests:** The authors declare no competing financial interests.

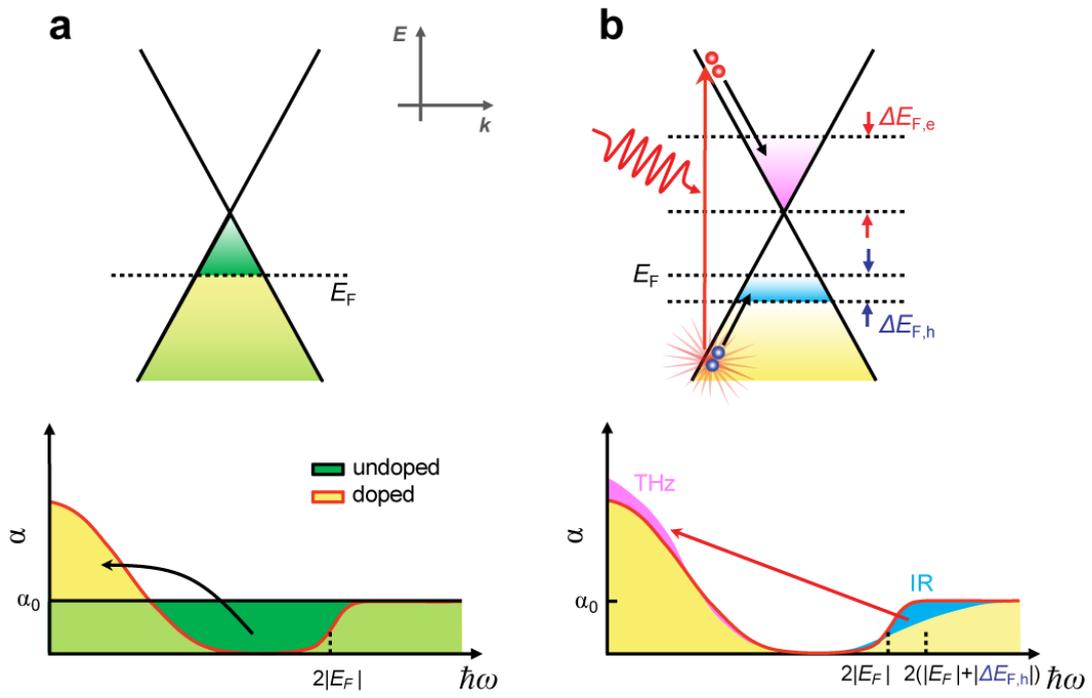

Fig. 1, J. Kim et al.

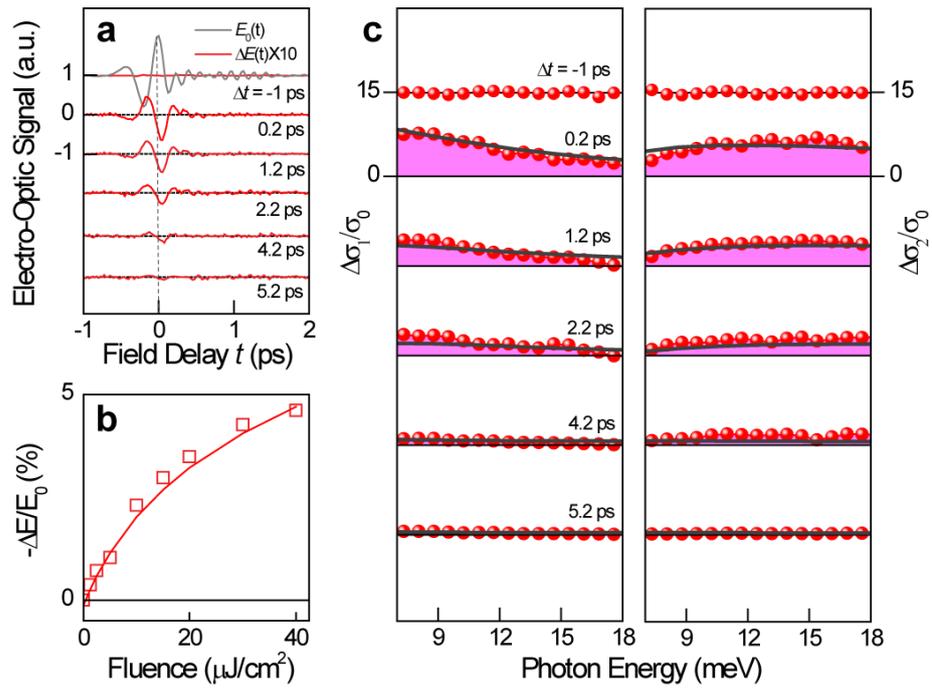

Fig. 2, J. Kim et al.

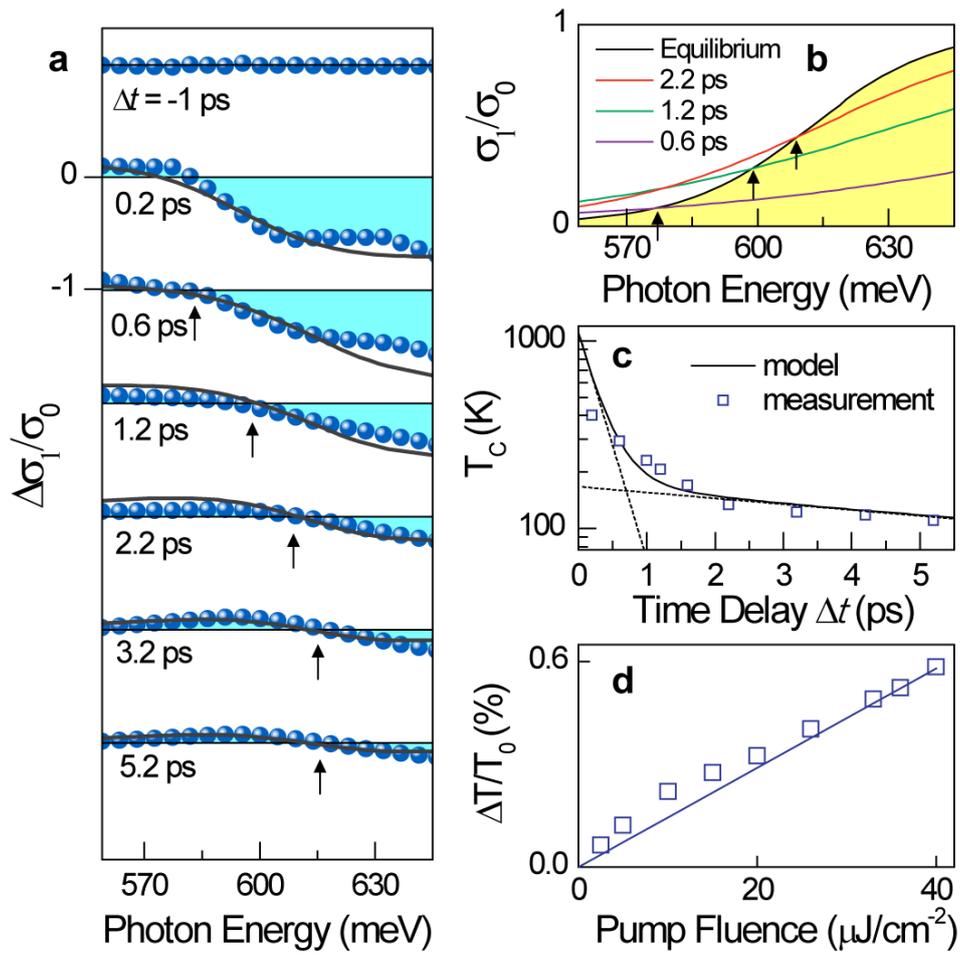

Fig. 3, J. Kim et al.

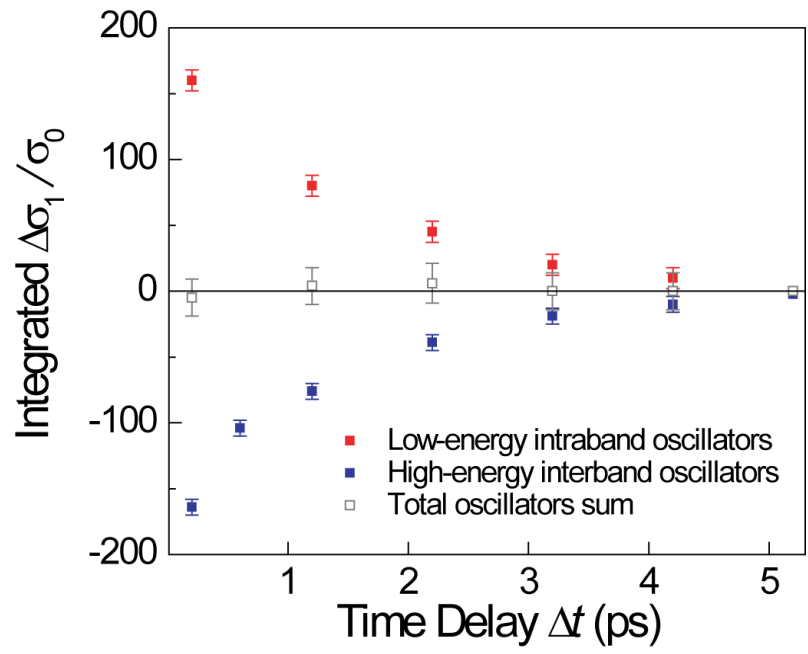

Fig. 4, J. Kim et al.